\documentclass[11pt]{article}
\usepackage[dvipdfmx]{color}
\usepackage{graphicx}
\usepackage{xcolor}
  \usepackage{amsthm,amsfonts}
  \usepackage{amsmath}
\usepackage{mathabx}
\usepackage{slashed}
\usepackage{ulem}
\usepackage{hyperref}
\usepackage{cleveref}
\usepackage{cite}

\usepackage{MnSymbol}%
\usepackage{wasysym}%

\newcommand{\bea}   {\begin{eqnarray}}
\newcommand{\eea}   {\end{eqnarray}}
\def\zzg{${\mathbb Z}_2\times{\mathbb Z}_2$-graded }
\def\zgg{${\mathbb Z}_2\times{\mathbb Z}_2$ }

\begin{document}
\renewcommand{\thefootnote}{\fnsymbol{footnote}}

\thispagestyle{empty}

\title{On the detectability of paraparticles beyond bosons and fermions}
\author{ Francesco Toppan\thanks{{E-mail: {\it toppan@cbpf.br}}}
\\
\\
}
\maketitle

{\centerline{
{\it CBPF, Rua Dr. Xavier Sigaud 150, Urca,}}\centerline{\it{
cep 22290-180, Rio de Janeiro (RJ), Brazil.}}
~\\
\maketitle

\begin{abstract}
In this paper I present the state of the art concerning the theoretical detectability
(and the open challenges for the experimental detectability) of a special class of paraparticles
beyond bosons and fermions. The particles under considerations, obeying a parastatistics,  are mutually exchanged via the permutation group and can exist in any space dimension (the anyons, which transform under the more general braid group, cannot exist in more than two space dimensions). 
\end{abstract}

\section{Introduction}	

A consistent formulation for parastatistics  (that is, the statistics obeyed by paraparticles beyond bosons and fermions) was presented in \cite{gre} (see also \cite{grme}) on the basis of the so called trilinear relations. In principle,
paraparticles could appear as  fundamental particles in Nature (e.g., entering some extensions of the standard model of particle physics) or emergent  (typical examples are the quasi-particles in condensed matter systems).  
\par
Two main classes of paraparticles can be considered.  Particles living in generic $D$ space dimensions can be exchanged via the permutation group, while in $D=2$ space dimensions (due to the special property that the first homotopy group of the punctured plane is nonvanishing), the braid group can play a role; for this group the braid exchange 
operator $B_{1\leftrightarrow 2}$ can admit the inequality $B_{1\leftrightarrow 2}^2\neq{\mathbb I}$, where ${\mathbb I}$ denotes the Identity operator. 
In this latter case a special class of emergent particles, constrained to live in two space dimensions, is given by the anyons (they received their name in \cite{wil}). \par
In this talk I present the state of the art concerning the theoretical detectability (and the open challenges for the experimental detectability) of the paraparticles, beyond bosons and fermions, of the first class;  unlike anyons these paraparticles can live in any space dimension $D$.\par
Before addressing properties and open questions of this class of parastatistics, it is instructive to briefly summarize the well-established features of two-dimensional anyons.\par
Anyons may be governed under exchange by one-dimensional or higher-dimensional representations of the braid group. The latter are now generally referred to as ``non-abelian anyons". The possibility of (abelian) anyons was first proposed in \cite{lemy} and then, from a different point of view, in \cite{gms81}. 
It was long believed that anyons could only be associated with one-dimensional unitary representations of the braid group, see \cite{wu}; the first suggestion that they  they could transform according to higher-dimensional representations of the braid group was presented in \cite{gms}. For a historical account on the prediction of the braid parastatistics satisfied by anyons one can see \cite{gol} and the references therein. Recently, anyonic particles have been experimentally detected, see \cite{expanyons}. \par
With this result the question concerning the existence of braid-exchanging anyonic particles is settled, both theoretically and experimentally. What about the other class of paraparticles, beyond bosons and fermions, which are exchanged via the permutation group? This talk presents the updated state of the art.\par
The paper is organized as follows: Section {\bf 2} recalls the traditional ``conventionality of parastatistics" arguments which state that paraparticles are not directly observables because they could always be recovered from ordinary bosons/fermions statistics. Unfortunately, prejudices were generated by accepting for granted the thesis without properly questioning the range of applicability of the underlying hypotheses;  this seriously harmed the development of certain disciplines, like the Rittenberg-Wyler color Lie (super)algebras \cite{{rw1},{rw2}} (discussed in Section {\bf 5}) whose serious investigations have been delayed by decades. Section {\bf 3} presents the very recent theoretical advancements which show how to overcome the assumptions leading to the conventionality of parastatistics (the two papers  \cite{{top1},{top2}}, appeared in 2021, were the first ones, as I will comment in the Conclusions, to point out the theoretical detectability of paraparticles based on the permutation group); Section {\bf 3} also outlines the possibility for experimental detection of these paraparticles. The \cite{maj} graded Hopf algebra framework for parastatistics is presented in Section {\bf4}; it is applied to  the construction, in a First Quantization approach, of multiparticle sectors. The ``$n$-bit" parastatistics induced by the 
\cite{{rw1},{rw2}} color Lie (super)algebras are presented in Sections {\bf 5} and, for a minimal case, {\bf 6}. It is quite rewarding, in a Conference on Group Theoretical Methods in Physics which rightly honors Hermann Weyl in many ways (the
``Weyl-Wigner  Award", the ``Weyl Prize") to point out that the minimal setting for detecting paraparticles is linked to  the logic behind a beautiful example conceived by Weyl in his classical book {\it Symmetry} \cite{wey}. This point is illustrated in Section {\bf 7}. Further considerations and comments on the detectability of paraparticles and an outline of the works in progress are presented in the Conclusions.

\section{``Conventionality of parastatistics": arguments and prejudices}

After the \cite{gre} proof that the parastatistics can be consistently introduced, a question naturally appeared. Why only ordinary bosons and fermions and not more general types of paraparticles have been observed in Nature?  This question led to speculations that, perhaps, paraparticles could not be directly observed. A mathematical result of \cite{ara} is often, but uncorrectly, cited in support of this thesis. It was proved in \cite{ara} that 
parafields can be reconstructed, via Klein operators, from ordinary fields. While this mathematical result is indeed true, it cannot be incorrectly applied to dismiss the existence of observable paraparticles. Clearly, something else is required; it is sufficient to point out that even ordinary, low-dimensional fermions can be reconstructed, through ``bosonization", from nonlocal bosonic operators \cite{col}. A very detailed and nuanced account of
various arguments put forward to dismiss the physical content of paraparticles is given in \cite{conventionality}; quite aptly,
the title of that paper is ``Conventionality of Paraparticles".   Quoting from the Introduction: ``It has often be claimed - sometimes offhand, sometimes with considerable supporting argument - that every theory of paraparticles is physically equivalent to some theory of regular bosons or fermions. For short, we'll simply call this the {\it equivalence thesis}".\par
The equivalence thesis is just a thesis that has to be proved under certain hypotheses.
 As recalled in \cite{conventionality}, none of the supporting arguments is fully satisfactory. The most stringent one is perhaps the
Doplicher-Roberts Reconstruction Theorem \cite{doro}, based on \cite{drharo} and the DHR no-go theorem \cite{dhr}.  The key ingredients are the implementation of superselection rules and a localization principle. The localization principle obviously applies to relativistic theory; it can even find application to nonrelativistic theory (that is, emergent paraparticles), provided that the creation operators are localized. \par
The sad story is that the equivalence thesis has been almost generally assumed for granted, disregarding the hypotheses under consideration and their range of applicability. In Section {\bf 5} the negative effects of this prejudice in the development of the physics based on the \zzg Lie (super)algebras will be discussed.

\section{Detectability of paraparticles: the present state of the art}

Before presenting more technical topics in the next Sections, I give here a concise summary  of the recent developments which challenged the equivalence thesis on the conventionality of parastatistics. The main methods to prove the theoretical detectability of paraparticles and the mechanisms to evade the \cite{doro} localization principle will be outlined. The papers which proved the theoretical detectability of paraparticles transforming under the permutation group are, in chronological order: \cite{top1}, \cite{top2}, \cite{waha}, \cite{nbits} (see also \cite{toptransm}). In all the above cases the strategy was the same, namely to prove that certain observables can produce results (eigenvalues) that cannot be recovered from ordinary bosons/fermions statistics. The following frameworks have been used: in \cite{{top1},{top2},{nbits},{toptransm}} a First Quantization of ${\mathbb Z}_2\times {\mathbb Z}_2$-graded paraparticles (the $2$-bit parastatistics that will be described in Section {\bf 5}). For the scope it is sufficient to investigate the $2$-particle sector. In \cite{top1}  observables are found which allow to discriminate \zzg parafermions  from their ordinary counterparts (in \cite{top2} the observables can discriminate the presence of  \zzg parabosons); in \cite{{nbits},{toptransm}} the presence of paraparticles can be directly read from the degeneracy of the multiparticle energy levels of matrix deformed quantum oscillators of de Alfaro-Fubini-Furlan \cite{dff} type. This is made possible because, contrary to \cite{{top1},{top2}}, the creation operators which produce excited states are not nilpotent. \par
In the \cite{{top1},{top2},{nbits},{toptransm}} works the \cite{doro} localization principle is evaded since the First Quantization does not require the notion of localization.\par
A different framework is presented in \cite{waha}, which shows that emergent paraparticles excitations appear in certain quantum spin Hamiltonians. The mechanism which allows in \cite{waha} to evade the localization principle is based on the fact that the excited states are created by non-localized operators which are of stringy nature.\par
The results of \cite{{top1},{top2},{waha},{nbits}} imply that the paraparticles are theoretically detectable at least in the realm of the emergent quasi-particles.\par
On the experimental side too there are pretty interesting developments. Perhaps the most significant ones 
are presented in \cite{parasim} (the simulation of paraparticles) and especially \cite{paraexp}. In this work the experimental group developed techniques to engineer paraparticles in the laboratory by using the spin of a trapped atomic ion and two of its bosonic modes of motion in the trap, with tailoring laser-induced couplings between them. The important message that one receives  is the possibility to manipulate paraparticles in the laboratory.  This does not mean that paraparticles transforming under the permutation group have already been experimentally detected.
The experimental detection of these paraparticles requires the observed results to be unequivocal; this means that they cannot be reproduced by ordinary bosons/fermions statistics. In short, what is needed is a signature of parastatistics by putting to experimental test those theoretical models (as the ones discussed in \cite{{top1},{top2},{waha},{nbits}}) which allow to do that. As far as I know, experimental tests of this type have not yet been performed. \par Further considerations on the theoretical and experimental detectability of paraparticles will be presented in the Conclusions. 
\section{The Hopf algebra formulation for parastatistics}

The \cite{{gre},{grme}} trilinear relations for parastatistics admit a nice realization in terms of the graded Jacobi identities of certain Lie superalgebras; this was pointed out in \cite{gapa} for parabosons and in \cite{pal} for both
parabosons and parafermions (unlike parabosons, a parafermion is a paraparticle which satisfies the Pauli exclusion principle and has fermionic-like properties). In the light of these results the theory of Lie superalgebras and their representations found application to parastatistics.\par
An alternative framework for parastatistics, more limited in scope, was introduced in \cite{maj} in terms of (graded) Hopf algebras endowed with a braided tensor product. This formalism is suitable for applications to the First Quantization of linear systems (as done in the papers \cite{{top1},{top2},{nbits}} mentioned above). The noninteracting multiparticle states are constructed, see \cite{{cachto},{cckt},{kuztop}},  from the Hopf algebra coproducts. The Hopf algebra framework cannot be applied to nonlinear dynamics; nevertheless, it can also be employed to integrable systems which, at least locally, can be linearized in terms of the action-angle variables.
The connection between the Hopf algebra's approach to parastatistics and the trilinear relations has been clarified in \cite{{anpo1},{kada1}}.\par
For our purposes the graded Hopf algebras under consideration are the Universal Enveloping Algebras  $ {\cal U}\equiv {\cal U}({\mathfrak{g}})$ of a graded Lie algebra
${\mathfrak g}$. The Hopf algebra ${\cal U}$ possesses:\\
- two algebraic structures, the associative multiplication $m$ and the unit $\eta$, where
\bea
m:{\cal U}\otimes {\cal U}\rightarrow {\cal U}, \quad (m: U_A\otimes U_B\mapsto U_A\cdot U_B),&\quad&
\eta: {\cal U}\rightarrow {\mathbb C}\quad (\eta: {\textrm{{\bf 1}}}\mapsto 1),
\eea
- two algebraic costructures, the coproduct $\Delta$ and the counit $\varepsilon$
\bea\label{costructures}
\Delta:{\cal U}\rightarrow {\cal U} \otimes  {\cal U}, \quad &
\varepsilon: {\mathbb C}\rightarrow {\cal U},
\eea
- an operation, the antipode $S$, relating structures and costructures,
\bea
S: {\cal U}\rightarrow {\cal U}.
\eea
The compatibility of structures and costructures is guaranteed by the properties:
\bea&\label{deltauu}
   \Delta(U_AU_B)=\Delta(U_A)\Delta(U_B),~~
  \varepsilon(U_AU_B)=\varepsilon(U_A)\varepsilon(U_B),~~
   S(U_AU_B)= S(U_B)S(U_A)&\nonumber
\eea
and
\bea\label{coassociativity}
   (\Delta\otimes id)\Delta(U)&=&(id\otimes \Delta)\Delta(U), \qquad \qquad \quad\nonumber\\
   (\varepsilon\otimes id)\Delta(U)&=&(id\otimes\varepsilon)\Delta(U)=U, \nonumber\\
   m(S\otimes id)\Delta(U)&=&m(id\otimes
   S)\Delta(U)=\varepsilon(U){\textrm{\bf{1}}}.
\eea
The first relation is the coassociativity of the coproduct.

The action on the identity {\bf 1} is
\bea\label{deltaid}
   &\Delta({\textrm{{\bf{1}}}})={\textrm{{\bf 1}}}\otimes{\textrm{{\bf{1}}}}, \qquad\qquad\quad~~~~
  \varepsilon({\textrm{{\bf{1}}}})=1,\qquad\qquad
   S({\textrm{{\bf 1}}})={\textrm{{\bf{1}}}}.&
\eea

The action on a primitive element, i.e. a generator $g\in {\mathfrak{g}}$, is
\bea\label{deltag}
   &\Delta({ g})={\textrm{{\bf 1}}}\otimes{g}+g\otimes {\textrm{{\bf 1}}}, \qquad\qquad
  \varepsilon({g})=0,\qquad\qquad
   S({g})={-g}.&
\eea
If we identify $g$ with, let's say, a Hamiltonian operator, the first relation on the left implies the $E_{1+2}=E_1+E_2$ additivity of the energy levels of the two-particle states.
Let $R$ be a representation of the Universal Enveloping Algebra ${\cal U}$ on a vector space $V$. The representation of the operators induced by the coproduct, denoted with a hat, gives
\bea\label{rep}
&R: {\cal U}\rightarrow V,\qquad {\widehat \Delta}:= \Delta|_R\in End (V\otimes V),\qquad 
 {\textrm{with}}\quad  {\widehat{ \Delta(U)}}\in V\otimes V. &
\eea
From the coassociativity of the coproduct one gets
$
{\widehat{ \Delta^{(n)}(U)}}\in V\otimes V\ldots \otimes V$ taken $n+1$ times.
This formalism allows to construct the $N$-particle Hilbert space ${\cal H}^{(N)}$ as a subset of the tensor products of $N$ single-particle Hilbert spaces ${\cal H}^{(1)}$. Let's set, for simplicity, ${\cal H}\equiv{\cal H}^{(1)}$. We have
\bea\label{tensorspaces}
{\cal H}^{(N)}&\subset &{\cal H}^{\otimes N}.
\eea
The \cite{maj} notion of the braided tensor product allows to introduce a nontrivial statistics. It is defined by the relation
\bea
(U_A\otimes U_B)(U_C\otimes U_D) &=& U_A\otimes {\textrm{\bf 1}}\cdot \Psi(U_B\otimes U_C)\cdot {\textrm{\bf 1}}\otimes U_D,
\eea
with $U_B$, $U_C$ braided by the ``braiding operator" $\Psi$ which satisfies a set of consistency conditions (see \cite{maj} for details).
In our applications to the $n$-bit parastatistics described in Section {\bf 5},  the braiding operator is just given by a set of signs:
\bea\label{epsilonbraiding}
&({\textrm{{\bf 1}}}\otimes U_B)(U_C\otimes {\textrm{{\bf 1}}}) =\Psi(U_B\otimes U_C)= (-1)^{\epsilon (B,C)} U_C\otimes U_B, 
\eea
with $\epsilon(B,C)=0,1. $

\section{Color Lie (super)algebras and $n$-bit physics}

Let us present a timeline of the introduction of (super)algebras in particle physics:
\\
~\\
$\bullet$ 1967: Coleman-Mandula presented \cite{coma} their famous no-go theorem which essentially states that
symmetries have a limited applicability in particle physics, because particles of different spins cannot be nontrivially accommodated in the same multiplet (their construction considered the symmetry groups, based on Lie algebras, which were available at that time). The introduction of superalgebras allowed to supersede the no-go theorem, leading to the
\\
$\bullet$ 1975: Haag-\L opusza\'nski-Sohnius classification \cite{haloso} of the super-Poincar\'e algebras based on point-particle Quantum Field Theory. This was not the end of the story,\\
$\bullet$ 1982: D'Auria-Fr\'e \cite{dafr} introduced the notion of extended supersymmetries (like the saturated $11$-dimensional $M$-algebra) which take into account the presence of extended objects (e.g., branes in string theory).\\
As a side story, which is not widely known,\\
$\bullet$ 1978: aimining at extending the \cite{kac} notion of ${\mathbb Z}_2$-graded  Lie superalgebras, Rittenberg-Wyler in \cite{{rw1},{rw2}} (see also \cite{sch}), introduced the ``color" Lie (super)algebras with extra gradings (the next simplest cases being the \zzg Lie algebras and the \zzg Lie superalgebras). The term ``color" was used for the envisaged applications to quarks.\\
~\\
These new color Lie (super)algebras structures  started being investigated by mathematicians, but received limited attention from physicists because they necessarily lead to parastatistics (then, why bother with paraparticles if their physics can always be recovered, as suggested by the {\it equivalence thesis}, from ordinary bosons/fermions?).
Even so, some interesting applications were made; in 1985 M. A. Vasiliev showed \cite{vas} that
some problems inherent to the construction of de Sitter supergravity are solved if one takes into account \zzg Lie superalgebras instead of the ordinary ${\mathbb Z}_2$-graded Lie superalgebras.\par
Recently, this situation changed with several advances from different directions. It was recognized, rather unexpectedly,  that \zzg color Lie superalgebras are dynamical symmetries of the nonrelativistic L\'evy-Leblond spinors \cite{{aktt1},{aktt2}}; the first example of a quantum mechanical system invariant under the one-dimensional  \zzg super-Poincar\'e algebra was presented in \cite{brdu}; systematic constructions of classical \cite{{akt1},{brusigma}} and quantum \cite{{akt2}} models were introduced;  a \zzg superspace was proposed \cite{{pon},{aido1},{aizt}}; \zzg parabosonic models were constructed \cite{kuto};  \zzg integrable systems introduced \cite{{bruSG},{z2z2int}}; a  bosonization approach was presented in \cite{que}. \par
The first implementation of a \zzg parastatistics was given in \cite{yaji}. Parastatistics induced by the trilinear relations based on the \zzg Jacobi identities were investigated in \cite{{tol1},{stvj1},{stvj2}}. The specific issue of the conventionality argument was not addressed. This question became relevant when Bruce-Duplij produced 
their \cite{brdu} model where a matrix quantum Hamiltonian is not only invariant under the one-dimensional \zzg super-Poincar\'e algebra, but it is also an example of supersymmetric quantum mechanics. Does the \zzg invariance produce physical measurable consequences? The answer is no for the single-particle quantum mechanical model of \cite{brdu}. In order to get a different physics  one has to introduce the multi-particle sectors recovered from First Quantization.

\subsection{From bosons/fermions $1$-bit physics to $n$-bit physics}

The ordinary physics based on bosons/fermions statistics can be described as a $1$-bit physics.  If bosons $B$ and fermions $F$ are building blocks represented by
\bea
B\equiv \Box, &\quad& F\equiv \blacksquare,
\eea
a composite $2$-particle state is a boson or a fermion according to
\bea
B+B\equiv ~(\Box\Box)\quad \Rightarrow\quad \Box,&&\nonumber\\
B+F\equiv~ (\Box\blacksquare)\quad \Rightarrow\quad \blacksquare,&&\nonumber\\
F+B\equiv ~(\blacksquare\Box) \quad\Rightarrow\quad \blacksquare,&&\nonumber\\
F+F\equiv ~(\blacksquare\blacksquare)\quad \Rightarrow\quad \Box.&&
\eea
The above  rule is reproduced by the ${\mathbb Z}_2$ addition
\bea\label{z2add}
 & 0+0=0, \quad 0+1=1,\quad 1+0=1,\quad 1+1=0 &
\eea
with the identification
\bea\label{identif}
B\equiv 0, &\quad & F\equiv 1.
\eea
The bosons/fermions statistics is defined by the following set of commutators/anticommutators obtained from the entries
of the table   
\bea\label{z2z2super}\label{onebittable}
\relax \begin{array}{|c|c|c|}\hline 
 &0&1 \\  \hline
0&[\cdot,\cdot]&[\cdot,\cdot]\\ \hline
1&[\cdot,\cdot]&\{\cdot,\cdot\} \\  \hline
\end{array}\qquad&\Rightarrow &\qquad \begin{array}{|c|c|c|}\hline 
 &0&1 \\  \hline
0&0&0\\ \hline
1&0&1 \\  \hline
\end{array},
\eea
where, in the right table, commutators are represented by $0$ and anticommutators by $1$; therefore the right table 
expresses the logical gate ``AND". \par
Many properties can already be inferred from these simple rules. For instance, while the spin-statistics connection (which implies that bosons have integer spin and fermions half-integer spin) requires the relativistic quantum field theory in order to be proved, already at this simple level something can be inferred. The composition of half-integer spins into irreducible representations is compatible with the ${\mathbb Z}_2$ addition (\ref{z2add}) given by the 
(\ref{identif}) identification:
\bea
\frac{1}{2}\otimes \frac{1}{2} &=& 0\oplus 1,\nonumber\\
F+F &=&B.\nonumber
\eea
One should note that the opposite identification (bosons as half-integer spin particles and fermions as integer spin particles) has to be ruled out at this level already. This implies that the nonrelativistic quantum mechanics  {\it does not require} a spin-statistics connection, but it is {\it compatible} with a spin-statistics connection {\it only under the proper identification of bosons/fermions with the integer/half-integer spin}. \par
As a matter of fact the simplest (minimal) parastatistics extensions of the bosons/fermions statistics are given by the $2$-bit physics induced by the Rittenberg-Wyler \zzg Lie (super)algebras. It is now time to present them and discuss their properties.\par

Color Lie (super)algebras can be defined \cite{nbits}  in terms of  ${\mathbb Z}_2^n$ gradings which induce $n$-bit parastatistics. For our purposes here it is sufficient to present the simplest  $n=2$ case which gives the ${\mathbb Z}_2^2\equiv {\mathbb Z}_2\times {\mathbb Z}_2$-graded Lie (super)algebras. By taking into account the $1$-bit, ${\mathbb Z}_2$-graded case, we end up with three inequivalent structures:
\bea\label{threecases}
{\textrm{{\it ~~~ i})}} && {\textrm{${\mathbb Z}_2$-graded Lie superalgebras,}} \nonumber\\
{\textrm{{\it ~~ ii})}} &&{\textrm{\zzg Lie superalgebras and}}\nonumber\\
{\textrm{{\it ~ iii})}} &&{\textrm{\zzg Lie algebras. }}
\eea
The corresponding graded Lie (super)algebra ${\mathfrak g}$ is decomposed into 
\bea\label{A1}
i)~~~~~&:&{\mathfrak g} = {\mathfrak g}_0\oplus {\mathfrak g}_1,\nonumber\\
ii) ~{\textrm{and}}~ iii)&:& {\mathfrak g} = {\mathfrak g}_{00}\oplus {\mathfrak g}_{01}\oplus {\mathfrak g}_{10}\oplus {\mathfrak g}_{11}.\eea
The even ($0$) and odd ($1$) generators in {\it i}) are bosonic (fermionic), while the four sectors of {\it ii}) and {\it iii}) are described by $2$ bits. The grading of a generator in {\it i}) is given by ${\vec \alpha}\equiv \alpha\in\{0,1\}$. The grading of a generator in
{\it ii}) and {\it iii}) is given by the pair ${\vec \alpha}=(\alpha_1,\alpha_2)$, with $\alpha_{1,2}\in\{0,1\}$.\par
Three respective inner products, with addition ${\textrm{mod}}~2$, are defined as:
\bea\label{innerproducts}
~ {i})~: ~~ {\vec\alpha}\cdot{\vec\beta} &:=& \alpha\beta\in \{0,1\},\nonumber\\
~ {ii})~: ~~ {\vec\alpha}\cdot{\vec\beta} &:=& \alpha_1\beta_1+\alpha_2\beta_2\in \{0,1\},\nonumber\\
~ {iii})~:~~  {\vec\alpha}\cdot{\vec\beta} &:=& \alpha_1\beta_2-\alpha_2\beta_1\in \{0,1\}.
\eea
The graded algebra ${\mathfrak g}$ is endowed with the operation $(\cdot,\cdot):{\mathfrak{g}}\times{\mathfrak g}\rightarrow{\mathfrak g}$. \par
Let $a,b,c\in{\mathfrak{g}}$ be three generators whose respective gradings
are ${\vec{\alpha}}, {\vec{\beta}},{\vec{\gamma}}$. The bracket $(\cdot,\cdot)$, defined as
\bea\label{roundbracket}
(a,b)&:=& ab -(-1)^{{\vec\alpha}\cdot{\vec \beta}}ba,
\eea
results in either commutators or anticommutators.\par
The operation satisfies the graded Jacobi identities
\bea
\label{gradedjac}
 (-1)^{\vec{\gamma}\cdot\vec{\alpha}}(a,(b,c))+
 (-1)^{\vec{\alpha}\cdot\vec{\beta}}(b,(c,a))+
 (-1)^{\vec{\beta}\cdot\vec{\gamma}}(c,(a,b))&=&0.
\eea

The grading $\deg[(a,b)]$ of the generator $(a,b)$ is the ${\textrm{mod}}~2$ sum
\bea\label{gradingcomp}
\deg[(a,b)]&=& {\vec{\alpha}}+{\vec{\beta}}.
\eea

The subcase $i$) in (\ref{innerproducts}) produces the (\ref{onebittable}) table of commutators/anticommutators.\par
The subcase $ii$) produces the \zzg Lie superalgebra table
\bea \label{2case}
&\begin{array}{c|cccc}&00&10&01&11\\ \hline 00&0&0&0&0\\10&0&1&0&1\\01&0&0&1&1\\11&0&1&1&0
\end{array}, &
\eea
The corresponding paraparticles are accommodated into $4$ sectors. The entries $0$ and $1$ respectively denote, as before,
a commutator or an anticommutator. The sectors $10$ and $01$ are fermionic in nature
due to the presence of $1$ in the diagonal entries (which implies a Pauli exclusion principle). On the other hand, unlike ordinary fermions, $10$ and $01$ particles mutually commute, producing a pair of \zzg parafermions.
The $00$ sector is purely bosonic, while the consistency of the construction requires the introduction of a $11$ sector which has a bosonic nature (due to the $0$ entry in the diagonal), but which anticommutes with the parafermions (the corresponding particles are usually called ``exotic bosons"). \par
The \zzg color Lie superalgebra is a true generalization of ordinary physics since, by leaving empty the $01$ and $11$ sectors, the $00$ and $10$ sectors alone reproduce the ordinary superalgebra of bosons/fermions.  As mentioned before, the prejudice against parastatistics prevented, since the very beginning, a systematic investigation of its induced physical  properties. \par
The subcase $iii$) produces the \zzg Lie algebra table 
\bea \label{3case}
&
\begin{array}{c|cccc}&00&10&01&11\\ \hline 00&0&0&0&0\\10&0&0&1&1\\01&0&1&0&1\\11&0&1&1&0
\end{array}. &
\eea
The corresponding paraparticles are also accommodated into $4$ sectors, but there are no fermionic-like particles obeying the Pauli exclusion principle since all diagonal entries are $0$ (i.e., commutators).  The $00$ sector is bosonic, while the three sectors $(10,01,11)$ define parabosonic particles which, for different sectors, mutually anticommute.
In the literature the term {\it color Lie superalgebra} is used to denote the presence of fermionic-like particles; the term
{\it color Lie algebra}  is used when no fermionic-like particles are present. \par

A minimal setting for realizing \zzg Lie (super)algebras consists in introducing the generators as ${\mathbb Z}_2^2$ graded $4\times 4$ matrices whose possible nonvanishing entries (denoted with the symbol ``$\ast$") are accommodated in each sector according to: 
{\footnotesize{\bea \label{z2z2matrix}
&M_{00} \equiv \left(\begin{array}{cccc} \ast&0&0&0\\0&\ast&0&0\\0&0&\ast&0\\0&0&0&\ast\end{array}\right), ~~
M_{10} \equiv \left(\begin{array}{cccc} 0&0&\ast&0\\0&0&0&\ast\\\ast&0&0&0\\0&\ast&0&0\end{array}\right),~~
M_{01} \equiv \left(\begin{array}{cccc} 0&\ast&0&0\\\ast&0&0&0\\0&0&0&\ast\\0&0&\ast&0\end{array}\right),~~
M_{11} \equiv \left(\begin{array}{cccc} 0&0&0&\ast\\0&0&\ast&0\\0&\ast&0&0\\\ast&0&0&0\end{array}\right).&\nonumber\\&&
\eea}} 
The entries can be real numbers, complex numbers or even differential operators. \par
We conclude by pointing out that the signs $(-1)^{{\vec\alpha}\cdot{\vec \beta}}$ entering (\ref{roundbracket}) appear as the signs (\ref{epsilonbraiding}) entering the braided tensor product.

\section{The minimal setting for parastatistics}

The fact that the minimal setting for a parastatistics is the $2$-bit physics makes it particular useful to perform the test which compares whether the results obtained from paraparticles can be reproduced by ordinary bosons/fermions.\par
A further simplification consists in investigating the ``minimal" \zzg Lie (super)algebras. They are the graded (super)algebras which admit one and only one generator in each graded sector. They have being classified in \cite{kuto}. For a minimal ${\mathbb Z}_2\times{\mathbb Z}_2$-graded Lie algebra the four generators $H$, $Q_i$ ($i=1,2,3$) 
can be assigned into each graded sector as
\bea
&H\in {\cal G}_{00},\quad Q_1\in{\cal G}_{10},\quad Q_2\in{\cal G}_{01},\quad Q_3\in{\cal G}_{11}.&
\eea
 The (anti)commutators defining the algebras and compatible with the gradings are
\bea\label{z2z2algstrcon}
\{Q_i,Q_j\}= d_k|\epsilon_{ijk}|Q_k, &\quad& [H,Q_i] = b_i Q_i,
\eea
where $\epsilon_{ijk}$ is the totally antisymmetric tensor normalized as $\epsilon_{123}=1$.
The six structure constants $d_i,b_i$ ($d_i,b_i\in{\mathbb K}$, with ${\mathbb K}={\mathbb R}~\textrm{or}~{\mathbb C}$) have to be constrained to satisfy the graded Jacobi identities. The resulting constraints are
\bea\label{z2z2lieconstraints}
&d_1(b_1-b_2-b_3)=d_2(b_2-b_3-b_1)=d_3(b_3-b_1-b_2)=0.&
\eea
In \cite{kuto} $8$ classes of inequivalent ${\mathbb Z}_2\times{\mathbb Z}_2$-graded Lie algebras are found (some of the classes depend on a sign or on real parameters). \par
Similarly,
the four generators $H$, $Q_1$, $Q_2$, $Z$ of the minimal ${\mathbb Z}_2\times{\mathbb Z}_2$-graded Lie superalgebras are assigned into each graded sector as
\bea
&H\in {\cal G}_{00},\quad Q_1\in{\cal G}_{10},\quad Q_2\in{\cal G}_{01},\quad Z\in{\cal G}_{11}.&
\eea
 The (anti)commutators defining these superalgebras are
\bea \label{z2z2superalgstrcon}\begin{array}{ccccccccccc}
\relax [H,Q_i]&=&a_iQ_i, && [H,Z]& =& bZ,&&[Q_1,Q_2]&=&cZ,\\
\{Q_i,Q_i\}&=&\alpha_iH, && \{Z,Q_i\} &=&\beta_i|\epsilon_{ij}|Q_j,&&&&\\
\end{array}&&
\eea
where the antisymmetric tensor $\epsilon_{ij}$ is normalized as $\epsilon_{12}=1$. After imposing the graded Jacobi identities $21$ classes of minimal ${\mathbb Z}_2\times{\mathbb Z}_2$-graded Lie superalgebras are encountered, see \cite{kuto}.
\par
At this point all ingredients are in place to analyze, see \cite{top2}, all possible parastatistics which are compatible with some given Hamiltonian such as the $4\times 4$ matrix oscillator
\bea\label{hamosc}
H_{osc}&=& {\footnotesize{\frac{1}{2}\left(
\begin{array}{cccc} -\partial_x^2+x^2-1&0&0&0\\0&-\partial_x^2+x^2-1&0&0\\0&0&-\partial_x^2+x^2+1&0\\0&0&0&-\partial_x^2+x^2+1
\end{array}
\right)}}
\eea
which possesses creation/annihilation oscillators $a^\dagger, a$ given by
\bea\label{aadagger}
a=\frac{i}{\sqrt{2}}(\partial_x+x)\cdot{\mathbb I}_4, && a^\dagger =\frac{i}{\sqrt{2}}(\partial_x-x)\cdot {\mathbb I}_4
\eea
and matrix raising (lowering) operators $f_{11}^\dagger, f_{10}^\dagger, f_{01}^\dagger$ ($f_{11}, f_{10}, f_{01}$) given by
\bea\label{fffmatrices}
&f_{11}^\dagger= {\footnotesize{\left(
\begin{array}{cccc} 0&0&0&0\\1&0&0&0\\0&0&0&0\\0&0&0&0
\end{array}
\right),}}\quad f_{10}^\dagger= {\footnotesize{\left(
\begin{array}{cccc} 0&0&0&0\\0&0&0&0\\1&0&0&0\\0&0&0&0
\end{array}
\right),}}\quad f_{01}^\dagger= {\footnotesize{\left(
\begin{array}{cccc} 0&0&0&0\\0&0&0&0\\0&0&0&0\\1&0&0&0
\end{array}
\right),}}&\nonumber\\
&f_{11}= {\footnotesize{\left(
\begin{array}{cccc} 0&1&0&0\\0&0&0&0\\0&0&0&0\\0&0&0&0
\end{array}
\right),}}\quad f_{10}= {\footnotesize{\left(
\begin{array}{cccc} 0&0&1&0\\0&0&0&0\\0&0&0&0\\0&0&0&0
\end{array}
\right),}}\quad f_{01}= {\footnotesize{\left(
\begin{array}{cccc} 0&0&0&1\\0&0&0&0\\0&0&0&0\\0&0&0&0
\end{array}
\right).}}&
\eea
The single-particle Hilbert space is spanned by $4$-component column vectors. The two-particle Hilbert spaces are
subspaces, see (\ref{tensorspaces}), of the Hilbert space spanned by $4^2=16$ column vectors. Depending on the different parastatistics assignments for $f_{10}^\dagger, f_{01}^\dagger, f_{11}^\dagger$, nine inequivalent $2$-particle quantizations are encountered, see \cite{top2} for details.
In particular, the bosonic and the \zzg parabosonic assignments produce different degeneracies of the energy levels with respect to the other 7 cases.\\
The finite-dimensional $2$-particle Hilbert spaces extracted from the bosonic and parabosonic assignments both contain $10$ states. $7$ of them are in common, while $3$ other states differ by a $\varepsilon$ sign ($\varepsilon=+1$ for bosons, $\varepsilon=-1$ for parabosons); let $v_i$ denotes the $16$-component column vector with $1$ in the $i$-th entry and $0$ otherwise, we have
\bea\begin{array}{lll}
U_{00,A} = v_1,\qquad &&\\
U_{00,B} = v_6,\qquad &U_{11}=\frac{1}{\sqrt 2}(v_2+v_5),\qquad&W_{11,\varepsilon} =\frac{1}{\sqrt 2}(v_{12}+\varepsilon v_{15}),\\
U_{00,C} = v_{11},\qquad&U_{10}=\frac{1}{\sqrt 2}(v_3+v_9),\qquad&W_{10,\varepsilon}=\frac{1}{\sqrt 2}(v_8+\varepsilon v_{14}), \\
U_{00,D} = v_{16},\qquad  &U_{01}=\frac{1}{\sqrt 2}(v_4+v_{13}),\qquad&W_{01,\varepsilon} =\frac{1}{\sqrt 2}(v_{7}+\varepsilon v_{10}),
\end{array}
\eea
where the suffix denotes the ${\mathbb Z}_2\times{\mathbb Z}_2$-grading of the vector in the parabosonic case.

In order to discriminate the \zzg parabosonic case from the bosonic case one should produce $2$-particle observables satisfying the following requirements:\\
~\\
{\it i}) they should apply to both bosonic and parabosonic Hilbert spaces,\\
{\it ii}) they should be hermitian, \\
{\it iii}) they should belong to the $00$-graded sector of the parabosonic theory in order to have real ($00$-graded) eigenvalues and\\
{\it iv}) they should produce the $\varepsilon $ eigenvalue when applied to the states $W_{11,\varepsilon}, W_{10,\varepsilon}, W_{01,\varepsilon}$.\par
~\par
The exchange matrices $X_{11}, X_{10}, X_{01}$ (the suffix indicates their respective ${\mathbb Z}_2\times{\mathbb Z}_2$-grading sector) mutually interchange the $11$, $10$ and $01$ sectors 
{\footnotesize{\bea\label{xxxmatrices}
&X_{11}= {\footnotesize{\left(
\begin{array}{cccc} 0&0&0&0\\0&0&0&0\\0&0&0&1\\0&0&1&0
\end{array}
\right),}}\quad X_{10}= {\footnotesize{\left(
\begin{array}{cccc} 0&0&0&0\\0&0&0&1\\0&0&0&0\\0&1&0&0
\end{array}
\right),}}\quad X_{01}= {\footnotesize{\left(
\begin{array}{cccc} 0&0&0&0\\0&0&1&0\\0&1&0&0\\0&0&0&0
\end{array}
\right).}}&
\eea}}
They are the building blocks for the construction of such observables.
~\par
By setting
\bea\label{xstu}
&X_s = X_{10}\otimes X_{10}, \quad X_t = X_{01}\otimes X_{01}, \quad X_u = X_{11}\otimes X_{11}, \quad X_\ast = X_s+X_t+X_u&\nonumber\\&&
\eea
one gets, e.g., for the $X_\ast$ observable which satisfies the above properties:
\bea
&X_{\ast} W_{11,\varepsilon}= \varepsilon W_{11,\varepsilon},\qquad  X_{\ast} W_{10,\varepsilon}= \varepsilon W_{10,\varepsilon},\qquad X_{\ast} W_{01,\varepsilon}= \varepsilon W_{01,\varepsilon}.&
\eea

\section{On Weyl and the minimal detection of paraparticles}

Before discussing the intepretation of the results presented in the previous Section and the key steps leading to the theoretical detectability of $2$-bit paraparticles in the minimal scenario, I will present a beautiful example concocted by Hermann Weyl in his popularizing book {\it Symmetry} \cite{wey}; it turns out to be relevant for this specific case of parastatistics.\par
Weyl presents the famous Clarke-Leibniz debate concerning the nature of the space (either absolute, thesis defended by the Newtonians, Clarke was one of them) or relative (thesis defended by Leibniz who anticipated some of the arguments later used by Mach). The Clarke-Leibniz debate was expressed in the metaphysical and theological language of the time. Weyl, in his book, reformulates the position of Leibniz (and also Kant) by expressing the relative nature of the space for the specific $Z_2$ parity transformation associated with the mirror symmetry. Weyl,
in his argument, mimicks the theological framework used by Clarke and Leibniz. \par
It is not surprising that Weyl, who introduced the notion of chiral/antichiral spinors, presented this argument; it is perhaps more surprising that Weyl's book was published in 1950, merely six years ({\it l'air du temps}) before Lee and Yang proposed \cite{leya} the parity violation in weak interactions (discovered immediately after by C. S. Wu and collaborators \cite{wuparity}).\par
The Weyl argument goes as follows. Let us suppose that God at the beginning of Time creates out of nothing, in the empty space, a hand. We have no way to say whether this hand is left or right
(in the empty space the hand is in all respect equivalent to its mirror image). Now, let us suppose that after
creating the first hand, God performs a second act of creation, producing a second hand. Only after the creation of the second hand the notion of right and left can be introduced. The second hand can be of the same type as
the first one (it can be aligned to the first hand via rotations and translations) or be of opposite type. Right-handedness or left-handedness is a relative notion
based on the referential provided by the first hand. Weyl uses this example to support the ``modern" Leibniz position that the space is relative. We can jokingly put it in this way: if we have Alice and Bob in different rooms, not directly seeing each other, Alice can wonder whether Bob is on the same side of the mirror as her.\par
~\par
Let's go back now to Weyl's two-hand scenarios. We can assign chirality $+1$ to what we call the right hand and chirality $-1$ to the opposite case, the left hand.\par
Put in front of the mirror, the total chirality of two opposite hands produces\bea
0=(\pm 1)+(\mp 1) &\Rightarrow &(\mp 1)+(\pm1) =0,
\eea
while the total chirality of two similar hands produces
\bea
\pm 2 = (\pm 1)+(\pm 1)&\Rightarrow& (\mp 1) +(\mp 1) = \mp 2.
\eea
In both cases the modulus is preserved (from the last equation $|+2|=|-2|=2$).\par
An empty space where two opposite hands are present is clearly distinct from an empty space where two similar hands
(no matter their absolute chirality) are present. This property has deep and paradoxical (paradoxical, not contradictory) consequences: despite the fact that a single right hand is equivalent to a single left hand, their combination is not. \par
~\par
I discussed in \cite{topchiral} the implications (well-known to the physicists working in the area)
for the construction of world-line supersymmetric sigma models. The irreducible representations of the one-dimensional ${\cal N}=4$-extended super-Poincar\'e algebra are given by multiplets of $4$ bosonic and $4$ fermionic time-dependent fields, whose transformations properties are induced by the quaternions. Similarly, the irreducible representations of the ${\cal N}=8$ extension are given by $8$ bosonic and $8$ fermionic fields
whose transformations properties are induced by the octonions, see \cite{krt}. The totally antisymmetric quaternionic
structure constant $\varepsilon_{ijk}$ of the three imaginary quaternions admits two (chiral/antichiral) normalizations:
\bea
&{\textrm{ either \quad $\varepsilon_{123}=+1$ \quad or \quad $\varepsilon_{123} = -1$.}}&
\eea
A worldline sigma model of a {\it single} ${\cal N}=4$ supermultiplet produces the same physics, irrespective of its chirality. This is not longer true for worldline sigma models constructed with two ${\cal N}=4$ supermultiplets.\par
Two ${\cal N}=4$ supermultiplets with $4+4$ component fields, if of {\it opposite} chirality, can be combined to produce an ${\cal N}=8$ supermultiplet with $8+8$ component fields.
This is not the case if the two ${\cal N}=4$ supermultiplets have the {\it same} chirality.\par
Conversely, taking an ${\cal N}=8$ supermultiplet and decomposing it with respect to an ${\cal N}=4$ subalgebra into two ${\cal N}=4$ irreducible supermultiplets, this necessarily produces two irreducible supermultiplets of opposite chirality.\par
One is tempting to paraphrase Orwell's {\it Animal Farm}: all unitarily equivalent representations are equal, but some are more equal than others.\par
~\par
What are the implications for the detection of paraparticles in the minimal scenario?
They are based on the following points:\par~
\\
$\bullet$ Let's have a source (natural or engineered in the laboratory)  of ordinary oscillators or paraoscillators which are created in a $2$-particle state. Let's select a state (via some measurement) which, as discussed in Section {\bf 6},
maps on this state the difference between oscillators/paraoscillators into a $\varepsilon =\pm 1$ ``chirality sign".\\
$\bullet $ Let us now apply a brand new detector (projector) which measures the $\pm 1$ ``chirality" in a yes/no experiment.\par~
\\
$\bullet$ Being brand new, the yes/no detector has to be calibrated. The first test (which corresponds to the Weyl's introduction of the first hand) is a  calibration test. At this stage one can invoke the ``conventionality of parastatistics" argument and establish that the result of the calibration corresponds to an ordinary $2$-particle oscillator.
\\
~\\
$\bullet$ After the first calibration  test is concluded, it is from the second test (corresponding to the introduction of the second hand) that the true measurements can start. One can assume, in an actual experiment, that the $2$-particle (para)oscillators are each time produced by manipulating some external parameters in a blind test.\\
~\\
$\bullet$ If the series of tests after the first one all produce the same output as the calibration test, one can conclude
that no paraoscillators are produced (the other possibility is that the detector is maybe defective!). If some of the tests produce a different output, one can reasonably conclude that this is evidence of a production of paraparticles. This is the parastatistics signature in this minimal setting.

\section{Conclusions}
This paper points out that paraparticles transforming under the permutation group are theoretically detectable,
at least as emergent quasi-particles in condensed matter systems or engineered in the laboratory with the methods described in \cite{paraexp}.  The minimal scenario for the detection of paraparticles was discussed at length. Non-minimal scenarios like the one presented in \cite{nbits} (the ``statistical transmutations of supersymmetric quantum mechanics") produce even more spectacular results since the detection of paraparticles can be directly read from the degeneracy of the energy eigenvalues in the multi-particle sector of systems like the de Alfaro-Fubini-Furlan \cite{dff} deformed oscillators. On the other hand, minimal scenarios are likely more ubiquous in Nature and, possibly, can be more easily
manipulated to conduct experiments in the laboratory.\par
The experimental detection of paraparticles requires theoreticians and experimentalists to meet halfway. The experimentalists need to test, in the laboratory, models which are known on theoretical grounds to produce a distinct signature of the parastatistics. \par
~\par
The first theoretical test to produce such a signature was presented in \cite{top1}.  
The addressed question was:  is it possible in a controlled system to check whether 
paraparticles could be reproduced by ordinary bosons/fermions (as  pretended by the {\it conventionality argument}) or not?
It was quite surprising  to realize that no such a test had been proposed before, 
despite the consistency of the parastatistics based on higher-dimensional representations of the symmetry group being well-established, see \cite{gold04}.
The possibility of such a test was offered by extending the Bruce-Duplij \cite{brdu} quantum model to a multi-particle sector. The test was then  reduced to a simple combinatorial analysis with a positive answer: \zzg  paraparticles (with parafermions) imply physically observable consequences.
The second paper \cite{top2} published in 2021 extended the test, with a positive answer, to \zzg parabosons.
Later, other types of models which pass similar theoretical tests were presented in \cite{waha} and in \cite{nbits}. 
So far, these four papers are the only ones introducing paraparticles which show a distinct signature of a parastatistics associated with representations of the symmetry group.\par

In order to facilitate a possible experimental detection of paraparticles,  we presented 
 in \cite{nbits} the $n$-bit parastatistics in terms of Boolean logic gates; they could help realizing the suitable experimental setups. In a forthcoming paper a blueprint to realize the minimal scenario for the detection of \zzg paraparticles will be detailed.\par
When I presented the state of the art on the detectability of paraparticles in the plenary talk at the ICGTMP Group33/35
Colloquium, G. A. Goldin suggested another possible scenario: searching for emergent non-abelian anyons which,
for suitable choices of the parameters and of the coupling constants, can transform under the permutation group instead of the more general braid group.\par 
~\par
A further scenario involves the detection of \zzg paraparticles from topological superconductors and insulators. The {\it periodic table of topological superconductors and insulators}, based on the so-called $10$-fold way, is a well-known tool in the classification of these materials, see \cite{{kit},{zir},{alzi},{rsfl}}. The $10$-fold way refers to the connection (see \cite{bae} for a  simple mathematical presentation) with the $10=3+7$ associative division and superdivision algebras. The $3$ associative division algebras are the well-known real numbers, complex numbers and quaternions.  $7$ extra ${\mathbb Z}_2$-graded superdivision algebras are added (producing the $10$-fold way). The elements of a ${\mathbb Z}_2$-graded superdivision algebra are split into even and odd sectors, where each sector is invertible. Furthermore, a superselection rule prevents introducing superpositions (linear combinations) of  generators of even and odd sectors.
The superselection prevents constructing divisors of zero. In \cite{aaca} we classified with Z. Kuznetsova the \zzg superdivision algebras over the reals, finding $13$ extra cases that have to be added to the $10$-fold way. We also pointed out that Hamiltonians constructed with parafermionic oscillators can be accommodated into the \zzg superdivision algebra setting (see \cite{aaca} for details). \par
~\par
An open question raised by the referee concerns the possible relation of \zzg parastatistics with higher-dimensional unitary representations of the permutation group. This issue requires a careful investigation. In the \cite{maj} framework the \zzg  braiding formula (\ref{epsilonbraiding}) is realized by unidimensional matrices (signs) rather than higher-dimensional matrices.  It is quite possible, but not guaranteed without a proper analysis, that an alternative description based on the Green's trilinear approach to parastatistics, see  \cite{gre} and \cite{{anpo1},{kada1}},
could be expressed in terms of higher-dimensional unitary reps of the permutation group. An interesting case which seems to support this possibility is provided by the braiding of Majorana qubits, see \cite{topmaj} and \cite{topmaj2}; a higher-dimensional braiding in the \cite{maj} framework  produce roots-of-unity truncations (labeled by an integer $s$) which implement Gentile-type \cite{gen} parastatistics. This is because at most $s-1$ paraparticles are accommodated into any multi-particle sector.
For any given $s$, the braided Majorana qubits belong to a representation of the braid group; in the $s\rightarrow \infty $ limit they are accommodated in a representation of the permutation group.  Interestingly enough, in this limit one recovers a Rittenberg-Wyler $Z_2\times Z_2$-graded parastatistics, see \cite{topmaj2}.
\par
~\par

To conclude, what about paraparticles as fundamental particles in Nature? This possibility cannot be excluded. In a work in progress with Rodrigo Rana we are showing that the \zzg parastatistics is compatible with the relativistic Quantum Field Theory. The key issue is the realization that the observables need to be restricted to the $00$-graded sector. A superselection rule which implies the consistency of the models is in place. One should also mention that
in \cite{spinstat} a spin-statistics connection is found for a class of noncommutative field theories; it has to be said that the \zgg grading induces a milder noncommutativity that the one presented in \cite{spinstat}.

\section*{Acknowledgments}

I am very pleased to thank  Naruhiko Aizawa and Zhanna Kuznetsova; we started together in a long collaboration the journey of understanding the properties of ${\mathbb Z}_2^n$-graded (aka, $n$-bit) physics.\\
I am indebted to the referee for historical remarks and for pointing out interesting open questions.\\
 This work was supported by CNPq (PQ grant 308846/2021-4).

\end{document}